\documentclass[prd,aps,groupedaddress,showpacs]{revtex4}
\usepackage{epsf,epsfig,graphicx}

\newcounter{muni}

\begin{document}
\hbadness=10000 \pagenumbering{arabic}

\title{Possible resolution of the $B\to \pi\pi$, $\pi K$ puzzles}

\author{Hsiang-nan Li$^{1}$}
\email{hnli@phys.sinica.edu.tw}
\author{Satoshi Mishima$^2$}
\email{satoshi.mishima@desy.de}

\affiliation{$^{1}$Institute of Physics, Academia Sinica, Taipei,
Taiwan 115, Republic of China,}

\affiliation{$^{1}$Department of Physics, Tsing-Hua University,
Hsinchu, Taiwan 300, Republic of China,}

\affiliation{$^{1}$Department of Physics, National Cheng-Kung
University, Tainan, Taiwan 701, Republic of China}

\affiliation{$^{2}$Theory Group, Deutsches Elektronen-Synchrotron
DESY, 22607 Hamburg, Germany}

\begin{abstract}

We show that there exist uncanceled soft divergences in the $k_T$
factorization for nonfactorizable amplitudes of two-body nonleptonic
$B$ meson decays, similar to those identified in hadron
hadroproduction. These divergences can be grouped into a soft factor
using the eikonal approximation, which is then treated as an
additional nonperturbative input in the perturbative QCD formalism.
Viewing the special role of the pion as a $q\bar q$ bound state and
as a pseudo Nambu-Goldstone boson, we postulate that the soft effect
associated with it is significant. This soft factor enhances the
nonfactorizable color-suppressed tree amplitudes, such that the
branching ratios $B(\pi^0\pi^0)$ and $B(\pi^0\rho^0)$ are increased
under the constraint of the $B(\rho^0\rho^0)$ data, the difference
between the direct CP asymmetries $A_{CP}(\pi^\mp K^\pm)$ and
$A_{CP}(\pi^0 K^\pm)$ is enlarged, and the mixing-induced CP
asymmetry $S_{\pi^0 K_S}$ is reduced. Namely, the known $\pi\pi$ and
$\pi K$ puzzles can be resolved simultaneously.

\end{abstract}

\pacs{13.25.Hw, 12.38.Bx, 12.39.St}

\maketitle

\section{INTRODUCTION}

The more precise data of the $B\to\pi\pi$, $\pi K$ decays have
sharpened the discrepancies with the theoretical predictions from
the factorization approaches, such as the perturbative QCD (PQCD)
approach based on the $k_T$ factorization theorem \cite{KLS,LUY}.
The observed $B^0\to\pi^0\pi^0$ branching ratio \cite{HFAG} remains
several times larger than the naive expectation. The direct CP
asymmetry of the $B^\pm\to\pi^0 K^\pm$ decays differs dramatically
from that of the $B^0\to\pi^\mp K^\pm$ decays. There is a deviation
between the extractions of the standard model parameter
$\sin(2\phi_1)$ from the penguin-dominated $B^0\to\pi^0 K_S$ modes
and from the tree-dominated $b\to c\bar cs$ modes. All these
discrepancies are closely related to the color-suppressed tree
amplitudes $C$ \cite{Charng2}. The $B^0\to\pi^0\rho^0$ branching
ratios from PQCD and QCD factorization (QCDF), being sensitive to
$C$, are also much lower than the data \cite{LY02,BN}. However, the
estimate of $C$ from PQCD is well consistent with the measured
$B^0\to\rho^0\rho^0$ branching ratio \cite{LM06}. Proposals
resorting to new physics \cite{FJPZ} mainly resolve the $\pi K$
puzzle without addressing the peculiar feature of $C$ in the
$\pi^0\pi^0$, $\pi^0\rho^0$, and $\rho^0\rho^0$ modes, while those
to QCD effects are usually strongly constrained by the $\rho\rho$
data \cite{BRY06}. It indicates the difficulty of resolving the
$\pi\pi$ and $\pi K$ puzzles simultaneously.

The color-suppressed tree amplitude $C$ seems to be an important but
the least understood quantity in $B$ meson decays. Viewing that all
the puzzles appear in the $C$-sensitive quantities, we shall
carefully investigate QCD effects on $C$, and their impact on the
$B\to\pi\pi$, $\pi K$ decays. Once a mechanism identified for $C$
respects the conventional factorization theorem, it is unlikely to
be a resolution due to the $B\to\rho\rho$ constraint mentioned above
\cite{LM06}. That is the reason the higher-order corrections
calculated in QCDF \cite{BY05}, which obey the collinear
factorization, cannot resolve the $\pi\pi$ puzzle. It has been
pointed out by Collins and Qiu \cite{CQ06} that the $k_T$
factorization breaks down in complicated QCD processes like
high-$p_T$ hadron hadroproduction because of the existence of soft
gluons in the Glauber region. To factorize the collinear gluons
associated with, say, one of the initial-state hadrons, one needs to
eikonalize the valence quark lines to which the collinear gluons
attach. Those eikonal lines, i.e., Wilson lines from another
initial-state hadron and the final-state hadrons, should cancel in
order to have the universality of the considered parton distribution
function. However, the required cancellation is not exact in the
$k_T$ factorization, though it is in the collinear factorization.
The $k_T$ factorization still holds for simple processes like deeply
inelastic scattering (DIS), which does not involve the Wilson lines
from the other hadrons. The Glauber gluons have been included as a mode
in the soft-collinear effective theory (SCET) recently \cite{BLO10}.

The above observation provides a clue for resolving the $\pi\pi$ and
$\pi K$ puzzles. It is easy to see that a factorizable amplitude,
involving only a $B$ meson transition form factor, mimics simple
DIS, and a nonfactorizable\footnote{Here a "nonfactorizable"
amplitude refers to a contribution that does not respect the naive
factorization assumption, which can also be called as a "spectator"
amplitude.} amplitude, involving dynamics of three hadrons, mimics
the complicated hadron hadroproduction. The $k_T$ factorization for
a factorizable $B$ meson decay amplitude has been proved
\cite{NL03}. The $k_T$ factorization for a nonfactorizable amplitude
has not, though it has been widely employed in the PQCD analysis.
Below we shall identify the residual infrared divergence in the
$k_T$ factorization for a nonfactorizable amplitude at one loop.
Contrary to high-$p_T$ hadron hadroproduction, this residual
infrared divergence can be factorized into a soft factor in two-body
nonleptonic $B$ meson decays, following the procedure in
\cite{CL09}, such that the universality of a $k_T$-dependent meson
wave function is restored. A nonfactorizable amplitude then remains
calculable in the PQCD approach after parameterizing the soft
factor. The color-suppressed tree amplitude $C$ receives a small
factorizable contribution, so the soft effect on a nonfactorizable
amplitude could be significant for $C$.

In Sec.~II we show the existence of residual infrared divergences
caused by Glauber gluons in a nonfactorizable emission diagram. It
is explained by means of contour deformation why Glauber gluons,
which do not meet the criteria of eikonalization in usual QCD
processes, can be factorized from two-body nonleptonic $B$ meson
decays. It is emphasized that the Glauber divergences do not appear
in the collinear factorization, such as the QCDF approach. In
Sec.~III we prove the factorization of the Glauber divergences into
a soft factor up to all orders, and derive its definition in terms
of nonlocal Wilson operators. We then investigate the numerical
impact of the soft factor on two-body nonleptonic $B$ meson decays
in Sec.~IV, and demonstrate that the $B\to\pi\pi$ and $\pi K$
puzzles mentioned above can be resolved. Section V contains the
conclusion.

\section{EIKONALIZATION OF GLAUBER GLUONS}

\begin{figure}[t]
\begin{center}
\begin{tabular}{cc}
\includegraphics[height=2.8cm]{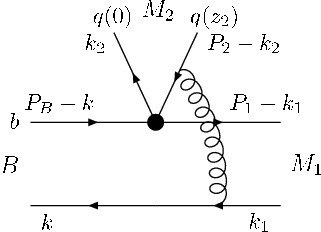}
\hspace{1mm} &
\includegraphics[height=2.5cm]{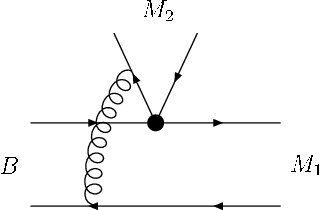}
\\
\hspace{-2mm}(a) & (b)
\end{tabular}
\caption{LO diagrams for a nonfactorizable amplitude.} \label{fig1}
\end{center}
\end{figure}

\begin{figure}[t]
\begin{center}
\begin{tabular}{ccc}
\includegraphics[height=2.5cm]{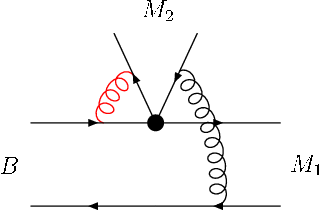}\hspace{0.3cm} &
\includegraphics[height=2.5cm]{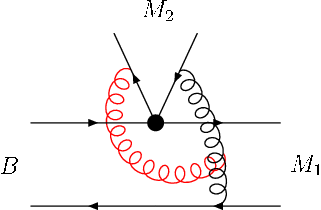}\hspace{0.3cm} &
\includegraphics[height=2.5cm]{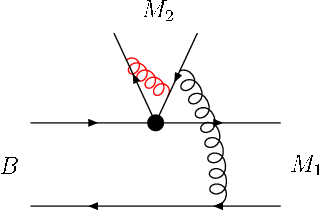}\hspace{0.3cm} \\
\hspace{-5mm} (a) &\hspace{-5mm} (b) &\hspace{-5mm} (c) \\
\\
\includegraphics[height=2.5cm]{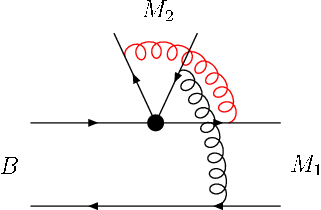}\hspace{0.3cm} &
\includegraphics[height=2.5cm]{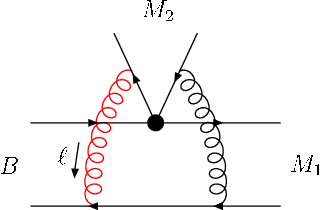}\hspace{0.3cm} &
\includegraphics[height=2.5cm]{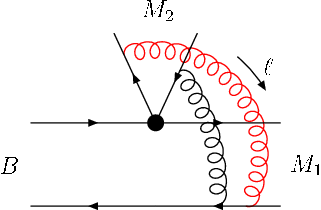}\hspace{0.3cm} \\
\hspace{-5mm} (d) &\hspace{-5mm} (e) &\hspace{-5mm} (f)
\end{tabular}
\caption{NLO diagrams for Fig.~\ref{fig1}(a)
that are relevant to the factorization of the $M_2$ meson wave
function.} \label{fig2}
\end{center}
\end{figure}

\begin{figure}[t]
\begin{center}
\includegraphics[height=2.5cm]{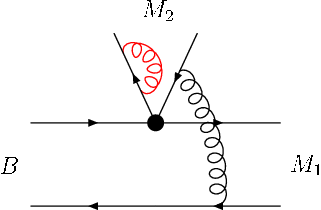}\hspace{0.3cm}
\includegraphics[height=2.5cm]{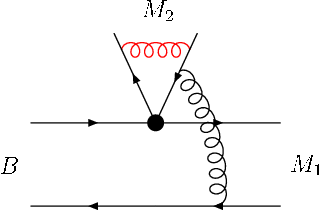}\hspace{0.3cm}

(a)\hspace{4.0cm}(b)

\caption{Two-particle reduced NLO diagrams for the $M_2$ meson wave
function.} \label{fig2r}
\end{center}
\end{figure}

Consider the $B(P_B)\to M_1(P_1)M_2(P_2)$ decay, where $P_{B,1,2}$
represent the momenta of the $B$, $M_1$, and $M_2$ mesons,
respectively. For convenience, we choose $P_1$ ($P_2$) in the plus
(minus) direction. Start with the leading-order (LO) nonfactorizable
emission diagram in Fig.~\ref{fig1}(a) resulting from the operator
$O_2$ \cite{REVIEW}, where the parton momenta $k$, $k_1$, and $k_2$
have been labelled. We add a radiative gluon of momentum $l$
collinear to $P_2$, which is emitted by the valence quark in $M_2$.
The attachment of the radiative gluon to the $b$ quark line shown in
Fig.~\ref{fig2}(a) leads to a Wilson line from infinity to the
origin, i.e, the weak vertex. This piece is factorized in color flow
by itself with the color factor $C_F$. The attachment to the hard
gluon in Fig.~\ref{fig2}(b) generates two Wilson lines, one of which
runs from the position $z_2$ of the valence anti-quark in $M_2$ to
infinity \cite{Li01}. The attachments to the virtual anti-quark in
Fig.~\ref{fig2}(c) and to the valence quark in the $M_1$ meson in
Fig.~\ref{fig2}(d) also generate the Wilson line running from $z_2$
to infinity. The combination of these three pieces with the same
Wilson line is factorized in color flow. As to the
next-to-leading-order (NLO) two-particle reducible diagrams, such as
the self-energy correction to the valence quark in
Fig.~\ref{fig2r}(a) and the gluon exchange between the valence quark
and the valence anti-quark in Fig.~\ref{fig2r}(b), their
factorization into the $M_2$ meson wave function is straightforward
\cite{Li01}.

The detail of the above treatment is similar to that
presented in \cite{NL03,Li01} for the pion form factor and the $B$ meson
transition form factor, which leads to the $k_T$-dependent $M_2$
meson wave function
\begin{eqnarray}
\Phi_{M_2}(x_2,k_{2T})&=&\int\frac{dz_2^+d^2z_{2T}}{(2\pi)^3
}\exp(-ix_2 P_2^-z_2^++i{\bf k}_{2T}\cdot {\bf z}_{2T})\nonumber\\
& &\times \langle
0|{\bar q}(z_2)\gamma_5\not\! n_+ W_+(z_2^+,{\bf z}_{2T};\infty)^\dagger
W_+(0,{\bf 0}_T;\infty)q(0)|M_2(P_2)\rangle, \label{pw1}
\end{eqnarray}
with the coordinate $z_2=(z_2^+,0,{\bf z}_{2T})$ of the valence
anti-quark and the dimensionless vector $n_+=(1,0,{\bf 0}_T)$ being
along the light cone. The path-ordered exponential $W_+$ collects
the Wilson lines mentioned above
\begin{eqnarray}
W_+(z^+,{\bf z}_T;\infty) = P \exp\left[-ig \int_0^\infty d\lambda
n_+\cdot A(z+\lambda n_+)\right]\;.
\end{eqnarray}
A vertical link to connect the two Wilson lines $W_+(z_2^+,{\bf
z}_{2T};\infty)^\dagger$ and $W_+(0,{\bf 0}_T;\infty)$ at infinity
is understood \cite{BJY,CKS10}.

\begin{figure}[t]
\begin{center}
\includegraphics[height=2.5cm]{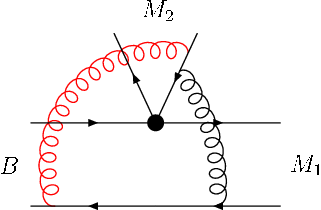}
\includegraphics[height=2.5cm]{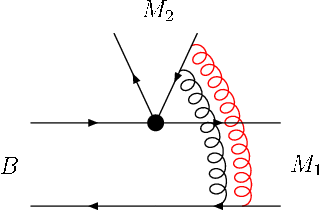}

(a)\hspace{3.5cm}(b) \caption{NLO diagrams that does not contribute
the Glauber divergence.} \label{fig2b}
\end{center}
\end{figure}

The other attachments shown in Figs.~\ref{fig2}(e) and
\ref{fig2}(f), and the second piece from Fig.~\ref{fig2}(b) should
cancel in order to have the universality of the $M_2$ meson wave
function in Eq.~(\ref{pw1}). We shall point out that it is not the
case, and the sum of the above three pieces gives a residual
infrared divergence. First, we justify the eikonalization of the
soft spectator in Figs.~\ref{fig2}(e), which demands the inclusion
of the NLO diagram in Fig.~\ref{fig2b}(a). Figure~\ref{fig2b}(a)
contains the four denominators
\begin{eqnarray}
[(P_2-k_2+l)^2+i\epsilon][(k-k_1+l)^2+i\epsilon]
[(k+l)^2+i\epsilon](l^2+i\epsilon),\label{4a}
\end{eqnarray}
with the loop momentum $l$, which define the following poles in the
$l^-$ plane
\begin{eqnarray}
& &l^-=-(P_2^--k_2^-)+\frac{|{\bf l}_T-{\bf k}_{2T}|^2}{2l^+}
-i\epsilon (+i\epsilon),\\
& &l^-=-k^-+\frac{|{\bf l}_T-{\bf k}_{1T}+{\bf
k}_T|^2}{2(l^+-k_1^++k^+)}+i\epsilon (+i\epsilon),\label{po1}\\
& &l^-=-k^-+\frac{|{\bf l}_T+{\bf k}_T|^2}{2(l^++k^+)}
-i\epsilon (-i\epsilon),\label{po2}\\
& &l^-=\frac{l_T^2}{2l^+}-i\epsilon (+i\epsilon),\label{pl1}
\end{eqnarray}
for the range of $0<l^+<k_1^+-k^+$ ($-k^+<l^+<0$). Here the
inequality $k_1^+>k^+$ has been assumed for convenience. The first
pole, being the furthest one, does not pinch the contour of $l^-$
actually. The two poles in Eqs.~(\ref{po1}) and (\ref{po2}) demand
that the contour goes through the region of $l^-\sim \Lambda_{\rm
QCD}$ for the soft spectator momentum $k^\mu\sim \Lambda_{\rm QCD}$
and the small transverse loop momentum $l_T\sim\Lambda_{\rm QCD}$,
$\Lambda_{\rm QCD}$ being the QCD scale. This observation does not
depend on the order of magnitude of the fourth pole. There is no
pinched singularity for $l^+>(k_1^+-k^+)$ and for $l^+<-k^+$,
because all the poles of $l^-$ are in the same half plane.

Figure~\ref{fig2}(e) contains the five denominators
\begin{eqnarray}
[(k_2+l)^2+i\epsilon][(P_2-k_2-k+k_1-l)^2+i\epsilon][(k-k_1+l)^2+i\epsilon]
[(k+l)^2+i\epsilon](l^2+i\epsilon).\label{2e}
\end{eqnarray}
Similarly, there is no pinched singularity for $l^+>(k_1^+-k^+)$ and
for $l^+<-k^+$. We consider the poles
\begin{eqnarray}
& &l^-=-k_2^-+\frac{|{\bf l}_T+{\bf k}_{2T}|^2}{2l^+}-i\epsilon(+i\epsilon),\label{pe1}\\
& &l^-=P_2^--k_2^--k^-+\frac{|{\bf l}_T+{\bf k}_{2T}-{\bf
k}_{1T}+{\bf k}_T|^2}{2(l^+-k_1^++k^+)}+i\epsilon(+i\epsilon),\label{po4}\\
& &l^-=-k^-+\frac{|{\bf l}_T-{\bf k}_{1T}+{\bf
k}_T|^2}{2(l^+-k_1^++k^+)}+i\epsilon(+i\epsilon),\label{po3}\\
& &l^-=-k^-+\frac{|{\bf l}_T+{\bf
k}_T|^2}{2(l^++k^+)}-i\epsilon(-i\epsilon),
\label{po5}\\
& &l^-=\frac{l_T^2}{2l^+}-i\epsilon (+i\epsilon),\label{pl2}
\end{eqnarray}
for $0<l^+<k_1^+-k^+$ ($-k^+<l^+<0$). The poles in Eqs.~(\ref{pe1})
and (\ref{po4}) are far from the origin by $l^-\sim O(m_B)$ due to
the large momenta $k_2$ and $P_2-k_2$, $m_B$ being the $B$ meson
mass, so they do not pinch the contour of $l^-$. The other three
poles in Eqs.~(\ref{po1}), (\ref{po2}) and (\ref{pl1}), identical to
those in Eqs.~(\ref{po4}), (\ref{po3}) and (\ref{pl2}),
respectively, for both ranges of $0<l^+<k_1^+-k^+$ and $-k^+<l^+<0$,
demand that the contour goes through the region of $l^-\sim
\Lambda_{\rm QCD}$.

We focus on the soft divergence from $l^+\to 0$ and $l_T\to 0$,
since the infrared finite piece contributes to the NLO hard kernel.
Picking up the poles of $O(\Lambda_{\rm QCD})$ in Eqs.~(\ref{po1})
and (\ref{po3}) for $0<l^+<k_1^+-k^+$, the $l^-$ dependence in
$(P_2-k_2+l)^2$, $(k_2+l)^2$, and $(P_2-k_2-k+k_1-l)^2$ is
negligible. Picking up the poles of $O(\Lambda_{\rm QCD})$ in
Eqs.~(\ref{po2}) and (\ref{po5}) for $-k^+<l^+<0$, the $l^-$
dependence is also negligible. Ignoring $k_{2T}^2\sim O(\Lambda_{\rm
QCD}^2)$ in $(P_2-k_2+l)^2$ and $(k_2+l)^2$, Figs.~\ref{fig2}(e) and
\ref{fig2b}(a) have the same amplitudes except a sign difference,
which is attributed to the emissions of the radiative gluon by the
valence quark and by the valence anti-quark in $M_2$. Because of
this soft cancellation in the pinched configuration, only the
$O(m_B)$ poles in Eqs.~(\ref{pe1}) and (\ref{po4}) for the range
$0<l^+<k_1^+-k^+$ are relevant. It implies that the contour of $l^-$
in Fig.~\ref{fig2}(e) can be deformed across the $O(\Lambda_{\rm
QCD})$ poles, and always remains at least of $O(m_B)$. That is, for a
gluon radiated from the energetic $M_2$ meson, we can consider only a
collinear divergence, instead of a soft divergence, if infrared
divergences are concerned. We then have the hierarchy
\begin{eqnarray}
k^+l^- \sim O(\Lambda_{\rm QCD}m_B)\gg |{\bf l}_T+{\bf k}_T|^2 \sim
O(\Lambda_{\rm QCD}^2),
\end{eqnarray}
for the denominator $(k+l)^2$. Therefore, the eikonal approximation
applies to the soft spectator on the $B$ meson side, giving the
propagator $1/(n_+\cdot l+i\epsilon)$.

The loop integral associated with Fig.~\ref{fig2}(e) is then written
as
\begin{eqnarray}
I_E&=&C_F\int \frac{d^4l}{(2\pi)^4}tr\bigg[...\frac{-i(\not\!\! P_2-\not
\! k_2-\not\! k+\not\! k_1-\not l)}{(P_2-k_2-k+k_1- l)^2+i\epsilon}
(-ig\gamma_\beta)\gamma_5\not\!\! P_2(-ig\gamma_\alpha)\frac{i(\not\!
k_2+\not
l)}{(k_2+l)^2+i\epsilon}\bigg]\nonumber\\
& &\times\frac{-i}{(k-k_1+l)^2+i\epsilon}
\frac{-gn_+^\alpha}{n_+\cdot
l+i\epsilon}\frac{-i}{l^2+i\epsilon},\label{il}
\end{eqnarray}
where the $...$ denotes the rest of the integrand, and
$\gamma_5\!\not\!\!P_2$ comes from the twist-2 structure of the
$M_2$ meson wave function. Combining Fig.~\ref{fig2}(f) and
Fig.~\ref{fig2b}(b) with the cancellation of the ordinary soft
divergences between them, we justify the eikonal approximation for
the spectator propagator $1/[(k_1-l)^2+i\epsilon]$, which gives
$1/(-n_+\cdot l+i\epsilon)$. The second piece from
Fig.~\ref{fig2}(b) contains the Wilson line running from the
position of the spectator in the $M_1$ meson to infinity, i.e., the
eikonal propagator $1/(-n_+\cdot l+i\epsilon)$. This piece with the
color factor $N_c/2$, together with Fig.~\ref{fig2}(f) with the
color factor $-1/(2N_c)$, leads to the loop integral the same as
Eq.~(\ref{il}) with the color factor $C_F$, but with
$1/(n_+\cdot l+i\epsilon)$ being replaced
by $1/(-n_+\cdot l+i\epsilon)$. Employing the principal-value
prescription
\begin{eqnarray}
\frac{1}{n_+\cdot l+i\epsilon}+\frac{1}{-n_+\cdot l+i\epsilon}=-2\pi
i\delta(l^-),\label{del}
\end{eqnarray}
we identify a NLO residual soft divergence from the Glauber region
with $l^-=0$, which seems to violate the universality of the $M_2$
meson wave function.

The spectator propagators in Fig.~\ref{fig2b} can also be replaced
by the Wilson line in the direction of $n_+$ for collecting the
Glauber divergences, if there are any. The eikonalization is
achieved by deforming the $l^-$ contour under the soft cancellation
observed above. We then examine the $l^+$ poles from the
denominators in Eq.~(\ref{4a}) with $l^-=0$ being demanded by
Eq.~(\ref{del}):
\begin{eqnarray}
& &l^+=\frac{|{\bf l}_T-{\bf k}_{2T}|^2}{2(P_2^--k_2^-)}
-i\epsilon,\\
& &l^+=k_1^+-k^++\frac{|{\bf l}_T-{\bf k}_{1T}+{\bf
k}_T|^2}{2k^-}-i\epsilon.
\end{eqnarray}
It is seen that both the $l^+$ poles are located in the lower half
plane, namely, Fig.~\ref{fig2b} does not contribute to the Glauber
divergences.
Figure~\ref{fig2b} generates only the ordinary soft divergences from
the region of the loop momentum
$l^\mu\equiv(l^+,l^-,l_T)\sim(\Lambda_{\rm QCD},\Lambda_{\rm QCD},
\Lambda_{\rm QCD})$ \cite{BT09}.
The effect from these ordinary soft gluons has been analyzed and
found to be negligible in two-body nonleptonic $B$ meson decays,
though it may be significant in $D$ meson decays \cite{LT98}.

The $l^+$ poles from Fig.~\ref{fig2}(e) in Eq.~(\ref{2e}) with
$l^-=0$ are given by
\begin{eqnarray}
& &l^+=\frac{|{\bf l}_T+{\bf k}_{2T}|^2}{2k_2^-}-i\epsilon,\label{plus1}\\
& &l^+=k_1^+-k^+-\frac{|{\bf l}_T+{\bf k}_{2T}-{\bf k}_{1T}+{\bf
k}_T|^2} {2(P_2^--k_2^--k^-)}+i\epsilon,\label{plus3}\\
& &l^+=k_1^+-k^++\frac{|{\bf l}_T-{\bf k}_{1T}+{\bf k}_{T}|^2}
{2k^-}-i\epsilon,\label{plus2}
\end{eqnarray}
in which only the first pole is of $O(\Lambda_{\rm QCD}^2/m_B)$. As
long as $k_1^+$ is of or greater than $O(\Lambda_{\rm QCD})$, we can
deform the contour of $l^+$, such that $l^+$ remains $O(\Lambda_{\rm
QCD})$, and the hierarchy
\begin{eqnarray}
k_2^-l^+ \sim O(m_B\Lambda_{\rm QCD})\gg |{\bf l}_T+{\bf k}_{2T}|^2
\sim O(\Lambda_{\rm QCD}^2)
\end{eqnarray}
holds. The valence quark carrying the momentum $k_2+l$ in
Eq.~(\ref{il}) can then be eikonalized into $n_{-\alpha}/(n_-\cdot
l+i\epsilon)$ with the vector $n_-=(0,1,{\bf 0}_T)$. The Glauber
divergence associated with Fig.~\ref{fig1}(a) is collected by
\begin{eqnarray}
I_a^{(1)}&=&g^2C_F\int \frac{d^4l}{(2\pi)^4}tr\bigg[...\frac{-i(\not
\!\! P_2-\not\! k_2-\not\! k+\not\! k_1-\not l)}{(P_2-k_2-k+k_1-
l)^2+i\epsilon}(-ig\gamma_\beta)\gamma_5\not\!\!
P_2\bigg]\nonumber\\
& &\times\frac{-i}{(k-k_1+l)^2+i\epsilon}\frac{1} {l^++i\epsilon}
\frac{-i}{-l_T^2+i\epsilon}2\pi i\delta(l^-),\label{vi0}
\end{eqnarray}
where the gluon propagator proportional to $1/l_T^2$ explicitly
indicates that the infrared divergence we have identified arises
from the Glauber region.

It is stressed that Eq.~(\ref{vi0}), derived from
Fig.~\ref{fig2}(e), contains the Glauber divergence associated with
Fig.~\ref{fig1}(b) as well: the left (right) gluon in
Fig.~\ref{fig2}(e) may become hard (soft) in some region of the loop
momentum $l$. We started with the eikonalization of the left gluon
in Fig.~\ref{fig2}(e), implying the attempt to isolate the Glauber
divergence associated with Fig.~\ref{fig1}(a). For consistency and
for avoiding double counting, we close the contour in the lower half
plane of $l^+$, and pick up only the pole $l^+=0-i\epsilon$ from the
eikonal propagator $1/l^+$, which corresponds to the $O(\Lambda_{\rm
QCD}^2/m_B)$ pole in Eq.~(\ref{plus1}). Another pole in
Eq.~(\ref{plus2}), corresponding to the on-shell right gluon,
contributes to the Glauber divergence associated with
Fig.~\ref{fig1}(b). Equation~(\ref{vi0}) is then simplified into
\begin{eqnarray}
I_a^{(1)} &\approx&
i\frac{\alpha_s}{\pi}C_F\int\frac{d^2l_T}{l_T^2}{\cal
M}_a^{(0)}({\bf l}_T), \label{vi}
\end{eqnarray}
where ${\cal M}_a^{(0)}$ denotes the LO amplitude from
Fig.~\ref{fig1}(a), and the imaginary logarithmic divergence is
explicit.

The Glauber divergence may not cause trouble, if the LO amplitude
${\cal M}^{(0)}$ is real. Expanding the decay width up to NLO, we
have
\begin{eqnarray}
|{\cal M}|^2 =|{\cal M}^{(0)}|^2+2{\rm Re}[{\cal M}^{(0)}{\cal
M}^{(1)*}].
\end{eqnarray}
According to Eq.~(\ref{vi}), the Glauber divergence will be purely
imaginary, if ${\cal M}^{(0)}$ is real, so it does not survive in
the second term ${\rm Re}[{\cal M}^{(0)}{\cal M}^{(1)*}]$. That is,
the Glauber divergence does not exist in the collinear
factorization. The absence of the Glauber divergence has been shown
up to two loops in the collinear factorization for hadron
hadroproduction \cite{CQ06}. On the contrary, ${\cal M}^{(0)}$ is
complex in the $k_T$ factorization, since partons carry transverse
momenta, and internal lines go on mass shell at finite momentum
fractions \cite{CLM08}. Thus the Glauber divergence contributes to
${\rm Re}[{\cal M}^{(0)}{\cal M}^{(1)*}]$ in the PQCD approach to
two-body nonleptonic $B$ meson decays. In the QCDF calculation
\cite{BY05} based on SCET
\cite{SCET}, the virtual anti-quark line in Fig.~\ref{fig2} has been
shrunk to a point, because this line is believed to be more off-shell than the hard
gluon. The pole in Eq.~(\ref{plus3}) then disappears, and the other
two in Eqs.~(\ref{plus1}) and (\ref{plus2}) are located in the lower
half-plane of $l^+$. As a consequence, the Glauber divergence seems not
to exist in the QCDF approach even at the amplitude level.

Below we discuss the absence of the Glauber divergence at the
amplitude level in QCDF in more details. The spin
structure associated with the $B$ meson wave function is written as
\cite{GN,KLS02}
\begin{eqnarray}
(\not\!\! P_B+m_B)\gamma_5 \left[\frac{\not\! n_+}{\sqrt{2}}\phi_B^{+}(k)
+\frac{\not\! n_-}{\sqrt{2}}\phi_B^{-}(k)\right] =-(\not\!\!
P_B+m_B)\gamma_5 \left[\phi_B(k)-\frac{\not\! n_+-\not\! n_-}{\sqrt{2}}
{\bar \phi}_B(k)\right], \label{bwp2}
\end{eqnarray}
with the functions
\begin{eqnarray}
\phi_B=\frac{1}{2}(\phi_B^{+}+\phi_B^{-})\;,\;\;\;
{\bar\phi}_B=\frac{1}{2}(\phi_B^{+}-\phi_B^{-})\;.
\end{eqnarray}
It has been known that only the structure $(\not\!\!
P_B+m_B)\gamma_5\not\! n_-$ contributes to the $B\to M_1$ transition
form factor, if choosing the $M_1$ meson momentum $P_1$ in the plus
direction. Assuming that the same structure contributes to the
nonfactorizable $B\to M_1M_2$ emission amplitude, the lower gluon
vertex in Fig.~\ref{fig1}(a) contains the matrix $\gamma^T$, since
it is sandwiched by $\not\! n_-\propto\gamma^+$ from the $B$ meson and
$\not\!\! P_1\propto\gamma^-$ from the $M_1$ meson. The upper gluon
vertex must contain the matrix $\gamma^T$ too. The Feynman rule
involving the anti-quark propagator in Fig.~\ref{fig1}(a) then
reduces to
\begin{eqnarray}
\frac{-i(\not\!\! P_2-\not\! k_2-\not\! k+\not\!
k_1)}{(P_2-k_2-k+k_1)^2+i\epsilon}(-ig\gamma^T)\gamma_5\not\!\!
P_2\approx\frac{-i k_1^+\gamma^-}
{2(P_2^--k_2^--k^-)k_1^++i\epsilon}(-ig\gamma^T)\gamma_5\not\!\! P_2,
\end{eqnarray}
in the collinear factorization for $\not\!\! P_2\propto\gamma^+$.
Cancelling $k_1^+$ in the numerator and in the denominator, this
anti-quark propagator is of $O(1/m_B)$, and can be shrunk to a point
in SCET. A similar argument applies to the NLO diagram Fig.~\ref{fig2}(e),
which leads to the Feynman rule
\begin{eqnarray}
\frac{-i(\not\!\! P_2-\not\! k_2-\not\! k+\not\! k_1-\not
l)}{(P_2-k_2-k+k_1-l)^2+i\epsilon}(-ig\gamma^T)\gamma_5\not\!\!
P_2\approx\frac{-i (k_1^+-l^+)\gamma^-}
{2(P_2^--k_2^--k^--l^-)(k_1^+-l^+)-l_T^2+i\epsilon}(-ig\gamma^T)\gamma_5\not\!\!
P_2.
\end{eqnarray}
The denominator becomes of $O(\Lambda_{\rm QCD}^2)$ as $l^+\to
k_1^+$ and $l_T\sim O(\Lambda_{\rm QCD})$. However, this infrared
region is suppressed by the numerator, so the anti-quark propagator does not
go on mass shell, and can be shrunk to a point. This explains
why the Glauber divergence does not appear in the QCDF calculation
of the nonfactorizable $B\to M_1M_2$ emission amplitudes.

If considering another spin structure $(\not\!\! P_B+m_B)\gamma_5\!\not\! n_+$
of the $B$ meson, the lower gluon vertex in
Fig.~\ref{fig1}(a) contains the matrix $\gamma^+$, and the upper one
contains $\gamma^-$. The Feynman rule involving the anti-quark
propagator in Fig.~\ref{fig1}(a) then becomes
\begin{eqnarray}
\frac{-i(\not\!\! P_2-\not\! k_2-\not\! k+\not\!
k_1)}{(P_2-k_2-k+k_1)^2+i\epsilon}(-ig\gamma^-)\gamma_5\not\!\!
P_2\approx\frac{-i (P_2^--k_2^-)\gamma^+}
{2(P_2^--k_2^-)k_1^++i\epsilon}(-ig\gamma^-)\gamma_5\not\!\! P_2,
\end{eqnarray}
which may go on mass shell as $k_1^+\to 0$. However, the virtual fermion
propagators differ by a minus sign in Figs.~\ref{fig1}(a) and \ref{fig1}(b),
since the hard gluon attaches to the anti-quark in $M_2$ in the former
and to the quark in the latter. Because of
the cancellation, we can neglect the spin structure
$(\not P_B+m_B)\gamma_5\!\not\! n_+$ at LO. At NLO, the Feynman rule for
Fig.~\ref{fig2}(e) with the structure $(\not\!\!
P_B+m_B)\gamma_5\!\not\!
n_+$ is given by
\begin{eqnarray}
\frac{-i(\not\!\! P_2-\not\! k_2-\not\! k+\not\! k_1-\not
l)}{(P_2-k_2-k+k_1-l)^2+i\epsilon}(-ig\gamma^-)\gamma_5\not\!\!
P_2\approx\frac{-i (P_2^--k_2^--l^-)\gamma^+}
{2(P_2^--k_2^--l^-)(k_1^+-l^+)-l_T^2+i\epsilon}(-ig\gamma^-)\gamma_5\not\!\!
P_2.
\end{eqnarray}
It implies that the anti-quark propagator diverges like
$1/\Lambda_{\rm QCD}^2$ as $l^+\to k_1^+$, which corresponds to the
pole in Eq.~(\ref{plus3}). The corresponding NLO correction to
Fig.~\ref{fig1}(b) is also Fig.~\ref{fig2}(e), but with the hard
gluon being on the left. In this case there is no
cancellation between the NLO correction to Fig.~\ref{fig1}(a) and
the NLO correction to Fig.~\ref{fig1}(b), and the region of $l^+\to k_1^+$
contributes to the Glauber divergence.
We postulate that the spin structure $(\not\!\! P_B+m_B)\gamma_5\!\not\! n_+$
should be kept, and the Glauber divergence exists at the
amplitude level in the QCDF calculation of the nonfactorizable emission
diagrams. Note that the structure $(\not\!\! P_B+m_B)\gamma_5$ on the
right-hand side of Eq.~(\ref{bwp2}) was adopted in the PQCD approach
(the contribution from the second wave function ${\bar\phi}_B$ is
power-suppressed \cite{KLS02}), and the virtual anti-quark line is
not shrunk to a point.

\section{SOFT FACTOR FROM GLAUBER GLUONS}

In this section we construct a soft factor $S({\bf b})$, ${\bf b}$
being the impact parameter conjugate to the transverse loop momentum
${\bf l}_T$, which collects the Glauber gluons to all orders. The
factorization at LO in the $b$ space is trivial:
\begin{eqnarray}
I_a^{(0)}=\int d^2b S^{(0)}({\bf b}){\cal M}_a^{(0)}({\bf b}),
\label{gl0}
\end{eqnarray}
where the function $I_a^{(0)}$ consists of the diagrams in
Fig.~\ref{fig1}, and the LO soft factor is simply the identity,
$S^{(0)}({\bf b})=1$. The $M_2$ meson wave function and
other subprocesses are contained in the nonfactorizable emission
amplitude ${\cal M}_a^{(0)}$. The factorization
of the soft factor at NLO has been explicitly demonstrated in
Sec.~II, and presented in Eq.~(\ref{vi}), which is expressed in the
$b$ space as
\begin{eqnarray}
I_a^{(1)}\approx\int d^2b S^{(1)}({\bf b}){\cal M}_a^{(0)}({\bf
b}),
\end{eqnarray}
with the NLO soft factor
\begin{eqnarray}
S^{(1)}({\bf
b})=i\frac{\alpha_s}{\pi}C_F\int\frac{d^2l_T}{l_T^2}e^{-i{\bf
l}_T\cdot {\bf b}}.
\end{eqnarray}
Adding the second radiative gluon emitted by the valence quark in
the $M_2$ meson, we have
\begin{eqnarray}
I_a^{(2)}\approx \frac{1}{2}\left(i\frac{\alpha_s}{\pi}C_F\right)^2
\int\frac{d^2l_{1T}d^2l_{2T}}{l_{1T}^2l_{2T}^2}{\cal M}_a^{(0)}({\bf
l}_{1T}+{\bf l}_{2T}), \label{vi2}
\end{eqnarray}
whose derivation is similar to the two-loop analysis in \cite{CQ06}.
The above factorization implies the next-to-next-to-leading-order
(NNLO) soft factor
\begin{eqnarray}
S^{(2)}({\bf
b})=\frac{1}{2}\left(i\frac{\alpha_s}{\pi}C_F\int\frac{d^2l_T}{l_T^2}e^{-i{\bf
l}_T\cdot {\bf b}}\right)^2.
\end{eqnarray}

Motivated by the above analysis up to NNLO, we postulate the
all-order definition of the soft factor,
\begin{eqnarray}
S({\bf b})=\langle 0|W_+(0,{\bf b};-\infty)W_+(0,{\bf
b};\infty)^{\dag} W_-(0,{\bf 0}_T;\infty)W_-(0,{\bf
0}_T;-\infty)^{\dag}|0\rangle,\label{soft}
\end{eqnarray}
where ${\bf b}$ can be interpreted as the transverse separation
between the weak decay vertex and the spectator. The link $W_-$
denotes another Wilson line operator
\begin{eqnarray}
W_-(z^-,{\bf z}_T;\infty) = P \exp\left[-ig \int_0^\infty d\lambda
n_-\cdot A(z+\lambda n_-)\right]\;,
\end{eqnarray}
with the coordinate $z=(0,z^-,{\bf z}_T)$. The net effect of the two
links $W_+(0,{\bf b};-\infty)W_+(0,{\bf b};\infty)^{\dag}$ demands
the vanishing of the component $l^-$ of the loop momentum, and the
off-shellness of a Glauber gluon by $l_T^2$ as indicated in
Eq.~(\ref{vi0}). We have included the additional link $W_-(0,{\bf
0};-\infty)^{\dag}$ in the above definition to demand the vanishing
of $l^+$, which plays a role similar to the anti-quark propagator in
$M_2$ for pinching the $l^+$ contour. It can be shown, by expanding
the Wilson line operators in the coupling constant, that
Eq.~(\ref{soft}) reproduces the NLO and NNLO soft factors presented
above.

We then extend the derivation of the soft factor to all orders by
means of induction \cite{NL03,Li01}, and demonstrate that it leads
to the operator definition in Eq.~(\ref{soft}). Assume that the
factorization holds up to $O(\alpha_s^N)$,
\begin{eqnarray}
G^{(j)}=\sum_{i=0}^{j} S^{(i)}\otimes {\cal M}_a^{(j-i)},\;\;\;\;
j=1,\cdots, N\;, \label{gbn}
\end{eqnarray}
where $\otimes$ represents the convolution in $b$,
\begin{eqnarray}
S^{(i)}\otimes {\cal M}_a^{(j-i)}\equiv\int d^2b S^{(i)}({\bf
b}){\cal M}_a^{(j-i)}({\bf b}).
\end{eqnarray}
In the above expression $S^{(i)}({\bf b})$ is given by the
$O(\alpha_s^{i})$ terms in the perturbative expansion of
Eq.~(\ref{soft}), and ${\cal M}_a^{(j-i)}({\bf b})$ stands for the
$O(\alpha_s^{j-i})$ nonfactorizable emission amplitude. We shall
show that the $O(\alpha_s^{N+1})$ diagrams $G^{(N+1)}$ can be
written as the convolution of the $O(\alpha_s^N)$ diagrams $G^{(N)}$
with the $O(\alpha_s)$ soft factor by employing the Ward identity,
\begin{eqnarray}
l_\mu G^\mu(l,k,k_1,k_2,\cdots)=0\;. \label{war}
\end{eqnarray}
In the above expression $G^\mu$ represents a physical amplitude with
an external gluon carrying the momentum $l$ and with $n$ external
quarks carrying the momenta $k$, $k_1$, $k_2$, $\cdots$. All these
external particles are supposed to be on mass shell in the
leading-power analysis here. It is known that factorization of a QCD
process in momentum, spin and color spaces requires summation of
many diagrams. With the Ward identity in Eq.~(\ref{war}), the diagram
summation can be handled in an elegant way.

Consider a complete set of $O(\alpha_s^{N+1})$ diagrams $G^{(N+1)}$
that are relevant to the factorization of the $M_2$ meson wave
function. Look for the gluon, one of whose ends attaches the outer
most vertex on the valence quark line in the $M_2$ meson. Let
$\alpha$ denote this outer most vertex, and $\beta$ denote the
attachments of the other end of the identified gluon. If $\beta$ is
located on the valence quark line in $M_2$, which corresponds to a
self-energy correction, and at the outer end of the valence
anti-quark line in $M_2$, which corresponds to a two-particle
reducible diagram, we have Fig.~\ref{fig2r} as the $O(\alpha_s)$
subdiagrams of $G^{(N+1)}$. In these two cases the identified gluon
can be factorized simply by inserting the Fierz transformation
\cite{Li01}, and absorbed into the $M_2$ meson wave function. If
$\beta$ attaches to the outer most vertex of the spectator line in
the $B$ meson, we eikonalize the anti-quark propagator adjacent to
$\beta$ into $-n_+^\beta/(n_+\cdot l+i\epsilon)$, which has appeared
in Eq.~(\ref{il}). Further eikonalizing the propagator of the
valence quark in $M_2$, we factorize a piece of contribution to the
NLO soft function
\begin{eqnarray}
g^2C_F\int\frac{d^4 l}{(2\pi)^4}\frac{n_-^\alpha}{n_-\cdot
l+i\epsilon}\frac{-ig_{\alpha\beta}}{l^2+i\epsilon}\frac{-n_+^\beta}{n_+\cdot
l+i\epsilon}G^{(N)}(l).\label{lsoft}
\end{eqnarray}
This piece can be obtained by contracting one gluon field in the
Wilson line $W_-(0,{\bf 0}_T;\infty)$ and another from $W_+(0,{\bf
b};-\infty)$.

For the other attachments of $\beta$ to lines in $G^{(N+1)}$, we
approximate the tensor $g_{\alpha\beta}$ in the propagator of the
identified gluon as \cite{Li01}
\begin{eqnarray}
g_{\alpha\beta}\approx\frac{-n_{+\alpha}l_\beta}{-n_+\cdot
l+i\epsilon}. \label{dec1}
\end{eqnarray}
The above approximation extracts the collinear enhancements
associated with the energetic $M_2$ meson, since the light-like
vector $n_{+\alpha}$ selects the minus component of $\gamma^\alpha$,
and the dominant component $l_{\beta=+}$ in the collinear region
selects the plus component of $\gamma^\beta$. The components
$l_{\beta=-,T}$ do not affect the collinear structure, because they
are negligible compared to the large momenta $P_1^+$ and $P_2^-$ of
$O(m_B)$. Equation~(\ref{dec1}) is applicable to the attachment to
the $b$ quark, which is free of collinear divergences. It is
certainly appropriate to adopt Eq.~(\ref{dec1}) for the attachments
to the internal lines and to the outer ends of the valence quark and
anti-quark lines in $M_1$, since it maintains the $l^-$ pole
structure of the propagators adjacent to the attachments. The above
observation can be checked by contracting $l_\beta$ to
Figs.~\ref{fig2}(b), \ref{fig2}(c), \ref{fig2}(d) and \ref{fig2}(f),
from which the conclusion in Sec.~II is drawn. The only attachment
of $\beta$, to which Eq.~(\ref{dec1}) does not apply, is the one to
the outer end of the spectator line in the $B$ meson, because of the
wrong location of the $l^-$ pole. This attachment has been handled
separately in Eq.~(\ref{lsoft}).

We have the Ward identity
\begin{eqnarray}
& &l_\beta\left\{G^{(N+1)\beta}_{\rm partial}+\left[\bar u(k_2)
\gamma^\beta\frac{1}{\not\! k_2-\not l}\cdots\right]
+\left[\cdots\frac{1}{\not l-\not\!\! P_2+\not\! k_2}\gamma^\beta
v(P_2-k_2)\right]\right.\nonumber\\
& &\left.+\left[\bar v(k_1)\gamma^\beta\frac{1}{-\not\! k_1-\not
l}\cdots\right]\right\}=0,\label{w2}
\end{eqnarray}
for it involves a full set of contractions of $l_\beta$ to all
lines in $G^{(N+1)}$. In the above expression the three diagrams
with $\beta$ being located at the outer ends of the valence quark
and anti-quark lines in $M_2$, and at the outer end of the spectator
line in the $B$ meson have been excluded from the set of
$G^{(N+1)}_{\rm partial}$. $u$ and $v$ are the spinors of a quark
and an anti-quark, respectively, and $\cdots$ represents the rest of
Feynman rules for $G^{(N)}$. Inserting the identities
\begin{eqnarray}
& &\bar u(k_2)\!\not l\frac{1}{\not\! k_2-\not
l}\cdots=-\bar u(k_2)\cdots,\nonumber\\
& &\cdots\frac{1}{\not l-\not\!\! P_2+\not\! k_2}\not l\,
v(P_2-k_2)=\cdots v(P_2-k_2),\nonumber\\
& &\bar v(k_1)\!\not l\frac{1}{-\not\! k_1-\not l}\cdots=-\bar
v(k_1)\cdots,
\end{eqnarray}
into Eq.~(\ref{w2}), we derive
\begin{eqnarray}
\frac{-n_{+\alpha}l_\beta}{-n_+\cdot l+i\epsilon}
G^{(N+1)\beta}_{\rm partial}= \frac{-n_{+\alpha}}{-n_+\cdot
l+i\epsilon}\left[\bar u(k_2)\cdots -\cdots v(P_2-k_2)\right]
+\frac{-n_{+\alpha}}{-n_+\cdot l+i\epsilon}\bar v(k_1)
\cdots.\label{w3}
\end{eqnarray}
The factor $-n_{+\alpha}/(-n_+\cdot l+i\epsilon)$ in the first term
on the right-hand side of Eq.~(\ref{w3}) contributes to the Wilson
lines in Eq.~(\ref{pw1}), which define the wave function for an
outgoing $M_2$ meson.

The second term on the right-hand side of Eq.~(\ref{w3}) contributes
to another piece of the NLO soft factor,
\begin{eqnarray}
g^2C_F\int\frac{d^4 l}{(2\pi)^4}\frac{n_-^\alpha}{n_-\cdot
l+i\epsilon}\frac{-i}{l^2+i\epsilon}\frac{-n_{+\alpha}}{-n_+\cdot
l+i\epsilon}G^{(N)}(l).\label{rsoft}
\end{eqnarray}
Combining Eqs.~(\ref{lsoft}) and (\ref{rsoft}), employing
Eq.~(\ref{del}), working out the integrations over $l^-$ and $l^+$,
and Fourier transforming the NLO soft factor into the $b$ space, we
derive
\begin{eqnarray}
G^{(N+1)}=S^{(1)}\otimes G^{(N)}+{\cal M}_a^{(N+1)}, \label{gbn1}
\end{eqnarray}
where ${\cal M}_a^{(N+1)}$ collects the $O(\alpha_s^{N+1})$
contribution that is free of the Glauber divergence. Applying the
same procedure to the operator definition in Eq.~(\ref{soft}), we
obtain the similar relation for the soft factor,
\begin{eqnarray}
S^{(i+1)}= S^{(1)}\otimes S^{(i)}, \label{gbs}
\end{eqnarray}
for $i=0$, 1, $\cdots$. At last, Eqs.~(\ref{gbn}), (\ref{gbn1}), and
(\ref{gbs}) lead to
\begin{eqnarray}
G^{(N+1)}=\sum_{i=1}^{N+1} S^{(i)}\otimes {\cal M}_a^{(N+1-i)}+{\cal
M}_a^{(N+1)}=\sum_{i=0}^{N+1} S^{(i)}\otimes {\cal M}_a^{(N+1-i)},
\label{gbn2}
\end{eqnarray}
with $S^{(0)}=1$. The above expression concludes the proof
for the factorization of the soft factor from the nonfactorizable
emission amplitude.

\section{IMPACT ON $B$ MESON DECAYS}

In this section we investigate the numerical effect of the soft
factor $S({\bf b})$. The soft factor has a dynamical origin similar to that of a
meson wave function: the former (latter) absorbs the Glauber
(collinear) gluons. The ${\bf b}$ dependence of $S({\bf b})$
can be obtained by nonperturbative methods or from experimental
data. For simplicity, we neglect this dependence, and parameterize
the soft factor associated with ${\cal M}_a^{(0)}$ as $\exp(iS_e)$,
\begin{eqnarray}
I_a\approx \exp(iS_e){\cal M}_a^{(0)},
\end{eqnarray}
where $S_e$ is treated as a real free parameter.
Glauber gluons emitted by the valence anti-quark of $M_2$ in the LO
diagram Fig.~\ref{fig1}(b) lead to
\begin{eqnarray}
I_b\approx \exp(-iS_e){\cal M}_b^{(0)},
\end{eqnarray}
where the minus sign is attributed to the radiation from the
anti-quark. In this case the right gluon in Fig.~\ref{fig2}(e) is
identified as the Glauber gluon. The above modified $k_T$
factorization formalism with the additional soft factor also applies
to the nonfactorizable emission amplitudes from other tree and penguin
operators. The soft factor for a nonfactorizable annihilation
amplitude is different, because Wilson lines in different directions are
involved. The study of this subject will be presented elsewhere. The
color-suppressed tree amplitude is small at LO due to the small
Wilson coefficient $a_2$ for the factorizable contribution and to
the pair cancellation between Figs.~\ref{fig1}(a) and \ref{fig1}(b)
for the nonfactorizable contribution. The presence of the soft
factor can convert the destructive interference in Fig.~\ref{fig1}
into a constructive one, resulting in strong enhancement. The soft
effect is expected to be minor in amplitudes other than
the color-suppressed tree, such as the color-allowed tree and
penguin (including annihilation), since they receive dominant
factorizable contributions.

We seek experimental constraints on the soft parameter $S_e$ by
comparing the data \cite{HFAG} (in units of $10^{-6}$)
\begin{eqnarray}
B(\pi^0\pi^0)&=&1.55\pm 0.19,\;\;
[(0.29^{+0.50}_{-0.20})]\nonumber\\
B(\pi^0\rho^0)&=&2.0\pm 0.5,\;\;[\approx 0.7]\nonumber\\
B(\rho^0\rho^0)&=&0.74^{+0.30}_{-0.27},\;\;
[(0.92^{+1.10}_{-0.56})],\label{da}
\end{eqnarray}
with the NLO PQCD predictions in the square brackets, which are
quoted from \cite{LMS05}, \cite{LM09}, and \cite{LM06},
respectively. The results from QCDF \cite{BN,VVQCDF} are similar.
The above comparison motivates us to postulate that the soft effect
is significant (negligible) in the decays with $M_2$ being a
pseudoscalar (vector) meson. That is, we associate a soft factor
with $M_2=\pi$, $K$, but not with $M_2=\rho$ (the soft effect
associated with the kaon is not crucial actually). A global fit to
the data of the $B\to VP$ decays based on flavor $SU(3)$ symmetry
also supported that the
color-suppressed tree amplitude is large (small), when $M_2$ is a
pseudoscalar (vector) meson \cite{CZ08}. Because the
$B^0\to\pi^0\rho^0$ decay involves both types of amplitudes with the
pion and the $\rho$ meson as $M_2$, it is natural that the
discrepancy is in between as indicated by Eq.~(\ref{da}). The larger
soft effect from the multi-parton states in the pion than in the
$\rho$ meson can be understood by means of the simultaneous role of
the pion as a $q\bar q$ bound state and as a Nambu-Goldstone (NG)
boson \cite{NS08}: the valence quark and anti-quark of the pion are
separated by a short distance, like those of the $\rho$ meson, in
order to reduce the confinement potential energy. The multi-parton
states of the pion spread over a huge space-time in order to meet
the role of a massless NG boson, which result in a strong Glauber
effect.

\begin{figure*}[t]
\begin{center}
\includegraphics[height=3.5cm]{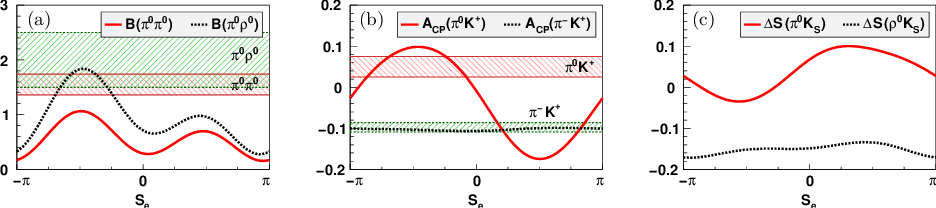}
\caption{$S_e$ dependence of (a) $B(\pi^0\pi^0)$ and
$B(\pi^0\rho^0)$ in units of $10^{-6}$, of (b) $A_{CP}(\pi^0 K^\pm)$
and $A_{CP}(\pi^\mp K^\pm)$, and of (c) $\Delta S_{\pi^0 K_S}$ and
$\Delta S_{\rho^0 K_S}$. The data (horizontal bands) for $\Delta S$
are not shown due to their large errors.} \label{fig3}
\end{center}
\end{figure*}

The factorization formulas for the $B\to\pi\pi$, $\pi\rho$, $\pi K$,
and $\rho K$ decays can be found in \cite{LMS05,LM062}. According to
our derivation, we multiply the $b$ quark nonfactorizable emission
amplitudes, both tree and penguin, by $e^{iS_e}$ ($e^{-iS_e}$) with
the hard gluon being emitted by the valence anti-quark (quark) in
$M_2$. The dependence on $S_e$ of those $C$-sensitive quantities is
displayed in Fig.~\ref{fig3}. The branching ratios $B(\pi^0\pi^0)$
and $B(\pi^0\rho^0)$ grow quickly with decreasing $S_e$ from the NLO
PQCD values in Eq.~(\ref{da}), and become close to the data when
$S_e$ reaches $-\pi/2$. Note that the Belle and BaBar data for
$B(\pi^0\pi^0)$ have different central values, $(1.1\pm 0.3\pm
0.1)\times 10^{-6}$ and $(1.83\pm 0.21\pm 0.13)\times 10^{-6}$,
respectively, and that our prediction is consistent with the
Belle's. The direct CP asymmetry $A_{CP}(\pi^0 K^\pm)$ increases
from the NLO PQCD result around $-0.01$ \cite{LMS05} to above $0.05$
for $S_e<-\pi/4$, whose agreement with the data $A_{CP}(\pi^0
K^\pm)=0.050\pm 0.025$ \cite{HFAG} is satisfactory. The deviation of
the mixing induced CP asymmetry $\Delta S_{\pi^0 K_S}\equiv S_{\pi^0
K_S}-S_{c\bar cs}$ descends from the NLO PQCD value $+0.07$ to
$-0.04$ for $S_e=-\pi/2$. Compared to the data $S_{\pi^0
K_S}=0.57\pm 0.17$ and $S_{c\bar cs}=0.672\pm 0.024$ \cite{HFAG},
the consistency has been improved. A measurement of sufficient
accuracy with an error of better than $\pm 0.04$ will be able to
verify the predicted shift in $S_{\pi^0 K_S}$ at the
$3\sigma$ level. The ratio $C/T=0.53e^{-2.2i}$
with $S_e\approx -\pi/2$ for the $B\to\pi\pi$ decays is close to the
extraction in \cite{Charng2}, $T$ being the color-allowed tree
amplitude. An equivalent viewpoint is that the $B(\pi^0\pi^0)$ data
constrain $S_e\sim -\pi/2$, which then leads to the predictions for
other quantities in Fig.~\ref{fig3}.

We have confirmed that $B(\pi^\mp\pi^\pm)$, $B(\pi^0\pi^\pm)$ and
all $B(\pi K)$ change slightly from those in \cite{LMS05}, since
they are less sensitive to $C$. $A_{CP}(\pi^\mp K^\pm)$ remains
around $-0.1$ \cite{KLS} for arbitrary $S_e$, and in agreement with
the data $A_{CP}(\pi^\mp K^\pm)=-0.098^{+0.012}_{-0.011}$
\cite{HFAG}. The small variation of the curve is attributed to the
soft effect on the nonfactorizable color-allowed tree and penguin
contributions. $S_{\rho^0 K_S}$ does not change much, because
$M_2=\rho$ in this case, and the involved $C$ is not modified. The
NLO PQCD prediction $\Delta S_{\rho^0 K_S}\approx -0.15$
\cite{LM062} is consistent with the data $S_{\rho^0
K_S}=0.63^{+0.17}_{-0.21}$ \cite{HFAG}. For those penguin-dominated
two-body modes without involving $C$, like $B\to\phi K$, their
mixing-induced CP asymmetries are not affected either.

To see the uniqueness of the pion, we investigate whether the
$B\to\eta'\pi^0$, $\eta'\eta'$ branching ratios exhibit a pattern
similar to that of the $B\to\rho^0\pi^0$, $\rho^0\rho^0$ ones. The
$\eta^{(\prime)}$ meson is unlikely to be a massless pseudo NG boson because
of the axial anomaly. As to $B(\eta\pi^0)$ and $B(\eta\eta)$, there
are only upper bounds for their data so far. The value
$B(\eta'\pi^0)=(1.2\pm 0.4)\times 10^{-6}$ \cite{HFAG} has been
measured, but the $B\to \eta'\eta'$ mode with
$B(\eta'\eta')=[1.0^{+0.8}_{-0.6}\pm 0.1 (< 2.4)]\times 10^{-6}$
\cite{BaBar} has not yet been seen. Both $B(\eta'\pi^0)$ and
$B(\eta'\eta')$ were predicted to be small in LO PQCD, roughly
$0.2\times 10^{-6}$ and $0.1\times 10^{-6}$ \cite{Xiao},
respectively. Namely, the PQCD prediction for the former is lower
than the data, but that for the latter might be reasonable, a
situation similar to the $B\to\rho^0\pi^0$, $\rho^0\rho^0$ case. We
also compare the patterns of the direct CP asymmetries in the $B\to
\eta^{(\prime)}K$ decays and in the $B\to \pi^0 K$ decays for the
same motivation. The data $A_{CP}(\eta' K^\pm)=0.016\pm 0.019$ and
$A_{CP}(\eta K^\pm)=-0.27\pm 0.09$ \cite{HFAG} are more or less in
agreement with the NLO PQCD results, $-0.06\pm 0.03$ and
$-0.12^{+0.14}_{-0.19}$ \cite{Xiao2}, respectively. The latter value
suffers from huge theoretical uncertainty, for the $B\to\eta
K^\pm$ decays involve cancellation of two amplitudes. It seems that
the $B\to \eta^{(\prime)}K$ modes behave normally, i.e., their
branching ratios and CP asymmetries coincide with the PQCD
predictions without the soft factor.

Our resolution differs from those based on new physics models, such
as the fourth-generation model \cite{HLMN,S08}, where it is the
electroweak penguin amplitude that is enhanced. These proposals, with
new weak phases being introduced, change $S_{\rho^0 K_S}$ and
$S_{\phi K_S}$. Our proposal differs from the elastic rescattering
models for final-state interaction, which involve multiple
intermediate states \cite{CHY}. A large $C$ has been generated
through the charge exchange mechanism in \cite{Chua08}. Since the
QCDF approach was employed there, the parameter scenario ``S4"
\cite{BN} or the inelastic scattering \cite{CCS} has to be
incorporated in order to get the correct result for $A_{CP}(\pi^\mp
K^\pm)$. The exchange mechanism, which requires turning of two
energetic quarks into opposite directions, is suppressed according
to the factorization theorem. Moreover, $\Delta S_{\pi^0K_S}$
remains positive in \cite{Chua08}.

\section{CONCLUSION}

In this paper we have identified the uncancelled Glauber divergences
in the $k_T$ factorization for the nonfactorizable $B$ meson decay
amplitudes, which are similar to those observed in hadron
hadroproduction. The divergences are factorizable and demand the
introduction of the soft factor, under which we have computed $C$ in
the PQCD approach, and found a possible simultaneous resolution of all the
puzzles: $B(\pi^0\pi^0)$ and $B(\pi^0\rho^0)$ are enhanced, the
difference between $A_{CP}(\pi^\mp K^\pm)$ and $A_{CP}(\pi^0 K^\pm)$
is enlarged, and $\Delta S_{\pi^0 K_S}$ is reduced for the single
soft parameter $S_e$ around $-\pi/2$. The constraint on $C$ from the
$B\to\rho\rho$ data is evaded, because of the special role of the
pion as a $q\bar q$ bound state and as a pseudo NG boson. Our
formalism involves some assumption, so an
evaluation of the ${\bf b}$ dependence of the soft factor, or even
of the soft parameter $S_e$ by nonperturbative methods will shed a
light on the resolution proposed here. The mechanism identified in
this work can be verified or falsified by more precise data in the
future. Based on our observation, we agree that the $B\to\pi K$ data
have not yet revealed a new physics signal \cite{M08}.


We thank S. Olsen and A. Soni for suggesting the
investigation of the $B\to\eta^{(\prime)}K$ and the
$B\to\eta^{(\prime)}\pi$ decays. We also thank M. Beneke
for his useful comments. This work was supported by the
National Center for Theoretical Sciences and National Science
Council of R.O.C. under Grant No. NSC-95-2112-M-050-MY3.

\end{document}